\documentclass[twocolumn]{webofc}

\usepackage[varg]{txfonts}   
\usepackage{hyperref}
\usepackage{url}
\usepackage[rightcaption]{sidecap}
\hypersetup{colorlinks=true,citecolor=blue,urlcolor=blue,linkcolor=blue}

\begin{document}
\title{The Stress-Force-Fabric relation across shear bands}

\author{\firstname{Carmen} \lastname{Lee}\inst{1}\fnsep\thanks{\email{cllee3@ncsu.edu}} \and
        \firstname{Emilien} \lastname{Azéma}\inst{2,3,4} \and
        \firstname{Karen E.} \lastname{Daniels}\inst{1}\fnsep\thanks{\email{kdaniel@ncsu.edu}}
}

\institute{Department of Physics, North Carolina State University, Raleigh, North Carolina, 27695, USA
\and
           LMGC, Universit\'e de Montpellier, CNRS, Montpellier, France 
\and
           Department of Civil, Geological, and Mining Engineering, Polytechnique Montréal, Montréal, Canada.
\and
           Institut Universitaire de France (IUF), Paris, France}

\abstract{The strength of granular materials is highly dependent on grain connectivity (fabric), force transmission, and frictional mobilization at the particle scale. Furthermore, these bulk properties are strongly dependent on the geometry and history of loading. It is well established that anisotropy in fabric and force transmission through a granular packing directly relates to the bulk scale strength of the packing via the Stress-Force-Fabric (SFF) relation. We have recently verified the validity of this framework for a broad variety of loading histories and geometries in experimental granular packings, using photoelastic disks to measure individual interparticle contact forces.  By tracking both particle positions and interparticle contact force vectors, we mapped the anisotropy of the fabric and forces to the macroscale stress and strain and found excellent agreement between the anisotropic particle-scale measures and the macroscale responses in experiments. Here, we present an analysis of the effect of strong spatial gradients (shear bands) using the SFF framework in a sheared annular geometry, finding that there are strong variations in contact orientation depending on the location within or outside the shear band, even though the principal loading direction is uniform. This highlights that the fabric connectivity significantly changes across the shear band but does not contribute to the direct loading of the material. We disentangle the effects of packing fraction gradients and boundary constraints on the differences in fabric orientation.
}
\maketitle
\section{Introduction \label{intro}} 

In the bulk, the mechanical behaviour of granular materials is strongly dependent on grain connectivity (fabric), force transmission, and frictional mobilization~\cite{jaeger_physics_1996}. Furthermore, these bulk properties are influenced by the geometry and history of loading. It is well established that anisotropy in fabric and force transmission through a granular packing directly relates to the bulk scale strength of the packing via the Stress-Force-Fabric (SFF) relation~\cite{rothenburg_analytical_1989, li_stressforcefabric_2013}.

Shear bands, which are mesoscale regions of localized plastic deformation, present a unique challenge in understanding granular materials~\cite{schall_shear_2010}. The strong spatial gradients in local strain and packing density compared to the bulk material make shear bands an important feature to study when trying to predict the behavior of materials under stress, particularly when loading exceeds the yield stress of the material.

We have recently verified the validity of the Stress-Force-Fabric framework for a broad variety of loading histories and geometries in experimental granular packings, using photoelastic disks to measure individual interparticle contact forces~\cite{lee_preprintA, lee_preprintB}. Here, we use the SFF relation to map emergent particle scale anisotropic structures to bulk friction in a shear granular material. We further address the spatial variation in the force and fabric distributions by comparing features in and outside of the shear band.

\section{Apparatus and Methods \label{apparatus}} 

\begin{figure*}
\centering
\includegraphics[width = \textwidth]{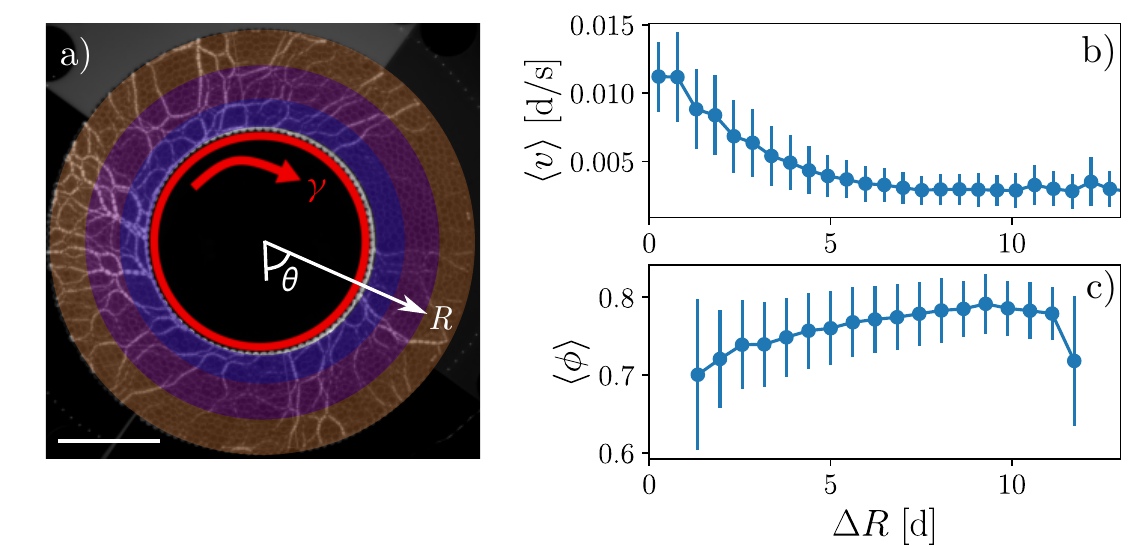}
\caption{(a) Top view of photoelastic particles viewed through a darkfield polariscope (white = higher stress) inside of an annular shear cell with inner boundary rotating at a constant strain rate $\dot\gamma$ and fixed outer boundary. Each particle provides a measurement of its own contact forces, which we measure relative to the radial polar coordinate $R$ and angular coordinate $\theta$. Overlaid on this image are coloured bands that correspond to the three radial bins used to examine the effect of strong spatial gradients in the annular shear cell and are referred to the inner, middle and outer bands. b) The average azimuthal velocity $v_\theta$ at different radial distances $\Delta R$ from the inner annulus boundary. Error bars correspond to the standard deviation in velocity. c) The average local packing fraction $\phi$ at different radial distances from the inner boundary. Error bars correspond to the standard deviation in $\phi$.}
\label{fig:1}       
\end{figure*}

In this study, we use a bidisperse mixture (diameters 9 mm and 11 mm, average $d = 10$~mm) of 1800 photoelastic disks ($N_s = 940, N_l = 860$)~\cite{abed_zadeh_enlightening_2019,daniels_photoelastic_2017} arranged in a single layer within an annular shear cell (Fig.~\ref{fig:1}(a)). The particles are contained in a dark field polariscope allowing for imaging of internal stress in the system, where lighter areas corresponds to higher stress. Particle positions, interparticle contacts and their resulting vector forces are extracted from the images using the Photoelastic Grain Solver algorithm~\cite{pegs2_photo-elastic_nodate}.

Slow shear is applied to the system by rotating the inner boundary at a rate of $0.015$ $d$/s 
and is rotated for full revolution, corresponding to a strain of  $\gamma =6$. The two boundaries of the annulus have particle-scale texture to prevent slip at the boundaries. We identify a shear band by plotting the azimuthal velocity profile $v$ at different radial distances from the inner wheel $\Delta R$ (Fig.~\ref{fig:1}(b)). Velocity profiles are averaged after the initial strain transient ($\gamma < 1$) is complete and the steady state is observed (see \cite{lee_preprintA}). We then split the annular data radially to encompass a strong shear band at the interior of the wheel ($0d< \Delta R < 4d$), the weak shear band ($4d< \Delta R < 8d$) and outside of the shear band ($8d< \Delta R < 12d$). These three regions have been notated in fig.~\ref{fig:1}(a) with colored bands. 

We calculate the local packing fraction $\phi$ by considering a 1.5$d$ circle surrounding the each particle. The local packing fraction corresponds to the ratio of area occupied by particles to the total area. Figure~\ref{fig:1}(c) shows the radial variation in packing fraction in steady state. We note that  the data is truncated near the boundaries as the local packing fraction calculation is sensitive to the presence of boundaries and misidentification of particles leading to higher experimental error. There is a less dense and more varied packing of particles within the shear band compared to outside of the shear band (middle and outer rings). 

\section{Stress-Force-Fabric Relation} 

To relate particle scale metrics to the bulk response, we use the Stress Force Fabric relation. We begin with defining a stress tensor in two dimensions using the micro-structural definition. We compute the two dimensional stress tensor $\sigma_{ij}$ via summation of the outer product of the force $\vec{f}$ and contact position vector $\vec{r}$ over the system of particles~\cite{bagi_microstructural_1999}:
\begin{equation}
    \sigma_{ij} = \frac{1}{S} \sum_{a \neq b} \vec{r^{ab}_i} \vec{f^{ab}_j},
    \label{eq:stresstensor}
\end{equation}
where in our two dimensional system, $S$ is the area of the annulus. Here we follow classical notation, where the contact position vector points from the center of the grain to the contact point, and the force vector points from the contact point in the direction of the contact force. From this formulation of the stress tensor, we calculate the bulk friction coefficient $\mu$,
\begin{equation}
    \mu = \frac{\tau}{P} = \frac{\sqrt{(\sigma_{11}-\sigma_{22})^2+4\sigma_{12}\sigma_{21}}}{(\sigma_{11} + \sigma_{22})/2},\label{eq:mu}
\end{equation}
where the numerator is the shear component of the stress tensor $\tau$, and the denominator is the mean pressure $P$.

\begin{SCfigure*}
\centering
\includegraphics[width=4.5in]{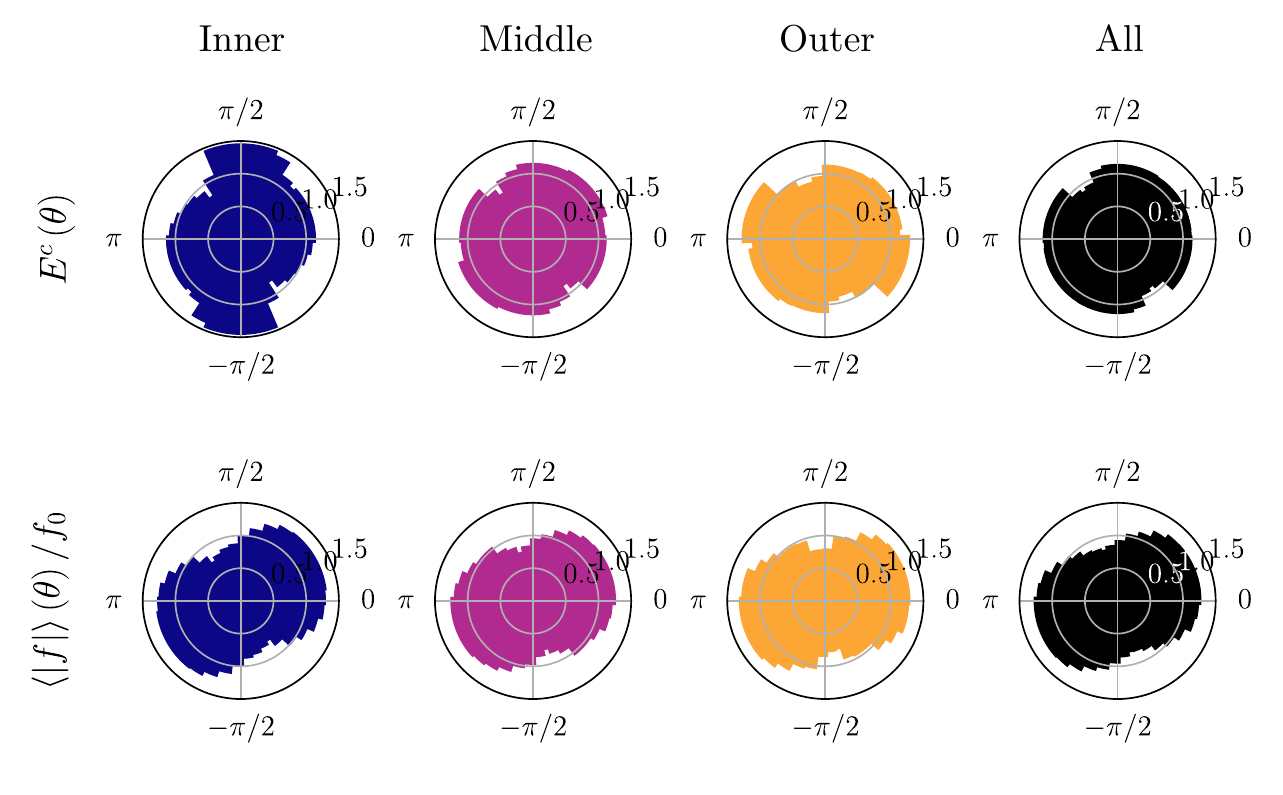}
\caption{Angular distributions of the contact density $E^c\left(\theta\right)$ (top row) and normalized force magnitude $\langle|f|\rangle/f_0$ for the three radial bands and the entire annulus. The normalizing force scale $f_0$ is the average normal force in the system. The angular distribution is calculated after the system comes to steady state $\gamma > 1$. The angular distribution correspond to the angle from the radial coordinate, with $\theta = 0$ and $\pi$ corresponding to the radial direction, where $\theta = \pi/2$ and $-\pi/2$ correspond to the azimuthal direction.}
\label{fig:2}     
\end{SCfigure*}

Along with bulk quantities, we also define three metrics that vary with contact angle $\theta$. The contact orientations, average normal forces, and average tangential forces in the granular material can be displayed as angular distributions fit to a Fourier expansion. The three distributions for the contact probability density $E^c\left(\theta\right)$, average normal force $\langle f_n \rangle (\theta)$ and average tangential force $\langle f_t \rangle (\theta)$ are 
\begin{subequations}
\begin{align}
E^c(\theta) & = \frac{1}{2\pi}\left[1+a \cos 2\left(\theta - \theta_a\right)\right], \label{eq:contacts}\\
\langle f_n \rangle (\theta) & = f_0\left[1+a_n \cos 2\left(\theta - \theta_a\right)\right], \label{eq:normals} \\
\langle f_t \rangle (\theta) & = f_0 a_t \sin 2\left(\theta - \theta_a\right) \label{eq:tangents},
\end{align}
\label{eq:expansions}
\end{subequations}
where the normalizing force scale $f_0$ is the average normal force over the entire ensemble, and each distribution has a principal angle $\theta_a$. The anisotropy of each distribution $a$, $a_n$ and $a_t$ is described by the amplitude of the leading order Fourier mode. Examples of the distributions are plotted in figure~\ref{fig:2} and are discussed fully in section~\ref{sec:distributions}.

We point the reader to the derivation presented in Rothenbug and Bathurst~\cite{rothenburg_analytical_1989} and Li and Yu~\cite{li_stressforcefabric_2013} to relate anisotropy in the distribution of the contacts, normal forces and tangential forces to the stress tensor resulting in a simple sum rule:
\begin{equation}
    \mu = \frac{1}{2}(a+a_n+a_t),
    \label{eq:SFF}
\end{equation}
showing that the bulk friction coefficient is closely tied to the amplitude of anisotropy, suggesting that the resistance to deformation in a granular material is fundamentally determined by the degree of anisotropy present within the granular assembly.

\section{Contact and force distribution \label{sec:distributions}} 

To examine the effect of shear banding in an annular geometry on the local fabric and force distribution within the SFF framework, we begin by plotting the angular distribution of contacts and forces. The top row of figure~\ref{fig:2} shows the angular distribution of the contact density $E^c\left(\theta\right)$ for each radial band and over all of the data. We observe that there is strong spatial variation from band to band, where the inner band contains contacts primarily aligned in the azimuthal direction, the middle band distribution of contacts is fairly isotropic, and the outer band contacts are strongly aligned in the approximately the radial direction. We note the likely effect of packing fraction, with increasing packing fraction outside of the shear band. In the transient strain regime, we see dilation in the shear band and compaction in the outer radial band. Migration of particles to the outer boundary happens early in the transient regime and results in a outward radial pressure. Particles outside of the shear band do not move following this initial motion and causes the radial fabric to remain frozen in the outer band.  The large variation in fabric is surprising, particularly in light of the nearly-constant force distribution that is shown in the bottom row of figure~\ref{fig:2}.

The bottom row of figure~\ref{fig:2} shows the averaged vector force magnitude for the contacts that are present at each angle. The distribution is normalized by the average normal force $f_0$ which we have set as the force scale for this dataset according to previous works~\cite{rothenburg_analytical_1989, lee_preprintA, lee_preprintB}. We neglect showing the tangential forces and normal forces separately here, as the tangential forces are small compared to the normal forces~\cite{lee_preprintA}. For each of the radial bands, the force alignment is strongest in the direction of shear approximately at approximately $\pi/4$. For increasing radial distance from the inner wheel, we note that there is a slight rotation of the principal loading direction toward the radial coordinate. This indicates that the strongest forces rotate toward the outer boundary in the outer band.  

The strong variation in principal contact direction in each band indicates that while there are many contacts between particles in the azimuthal direction (inner band) and radial direction (outer band), these contacts have a small force magnitude and are not responsible for bearing the majority of the stress. The discrepancy between the contact distribution and the force distribution shows that while the fabric connectivity is highly dependent on the radial position, the direct loading is not.

\section{Bulk friction \label{sec:mu}} 

\begin{SCfigure*}\includegraphics[width=5in]{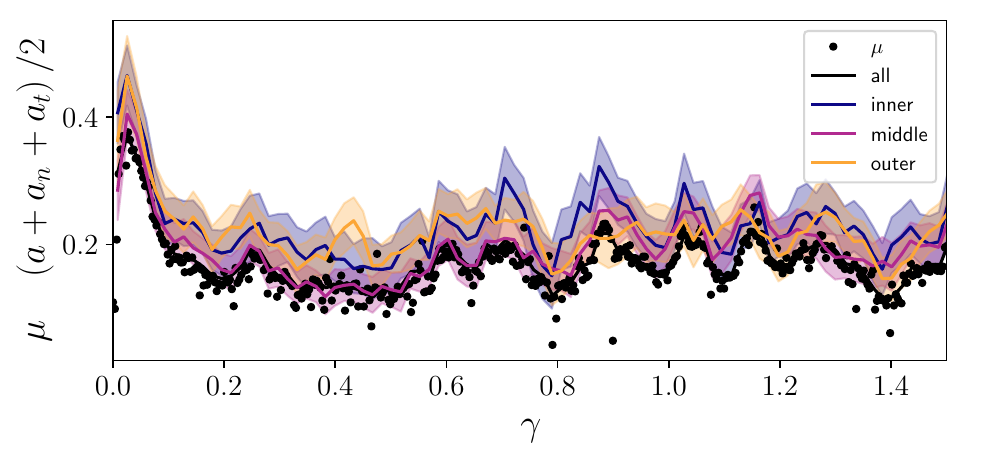}
\caption{Comparison of the bulk friction coefficient $\mu$ calculated from the stress tensor to the prediction from the SFF relation, as a function of strain $\gamma$. Each colour represents the calculated values for the inner, middle, and outer radial bands, with the bars indicating 1 standard deviation in the fits. }
\label{fig:3}      
\end{SCfigure*}

Finally, to compare difference between the radial positions and the bulk properties of the annular We begin cell, we plot the bulk friction coefficient $\mu$ as it develops with shear $\gamma$ (Figure~\ref{fig:3}). We begin by plotting the bulk friction coefficient calculated from the microstructural definition of the stress tensor over the entire granular material as presented in equation~\ref{eq:mu} (black dots). Separately, we calculate the bulk friction predicted by the SFF relation using the half-sum of the anisotropy measures in the contacts $a$, normal forces $a_n$, and tangential forces $a_t$ (eq.~\ref{eq:SFF}). Represented with solid lines, we perform this calculation for the entire annulus (black line) and each radial band. As well, we show the standard deviation in this calculation represented by colored bands. Although we calculated data over an entire strain window up to $\gamma =6$, we present a sample of these data to best illustrate the initial transient regime along with a section of the steady-state strain.

The two methods of calculating the bulk friction coefficient over the entire granular assembly match very well over the entire strain window. Each of the radial band calculations show that generally, $(a+a_n+a_t)/2$ in each band overestimates the bulk friction. Because the contact distributions vary radially, each radial band presents a higher anisotropy than in the bulk and leads to an elevated $a$ and in turn elevates the predicted bulk friction coefficient.

While the strongly-sheared inner band follows the dynamics of the entire granular material, we tend to see it exaggerates the overall behaviour. The outer band on the other hand, follows the dynamic behaviour in $\mu$ when measured over all of the  particles less closely, indicating that the majority of the response to stress happens most strongly in the shear band. Interestingly, the middle band matches the bulk values within the transient regime up until $\gamma \sim 0.8$, likely as the result of being an average response between the two boundaries while the shear band is being developed. We note that the bulk friction coefficient in the annular shear cell is expected to vary with radial position and we point the reader to work by Fazelpour and Daniels~\cite{fazelpour_controlling_2023} on measuring non-local rheology and $\mu(I)$-rheology for a detailed study on the change in shear stress and pressure with changing radial position in the annular shear cell.

In this work, we have presented an analysis of the effect of strong spatial gradients (shear bands) using the SFF framework in a sheared annular geometry. There are strong variations in contact orientation depending on the location within or outside the shear band, even though the principal loading direction is uniform. This highlights that the fabric connectivity significantly changes across the shear band but does not contribute to the direct loading of the material. We also show the importance of considering the entire granular material when using anisotropy measures to predict bulk properties like the bulk friction coefficient.

\end{document}